\documentclass[twocolumn,showpacs,preprintnumbers,amsmath,amssymb]{revtex4}

\usepackage{epsfig}
\usepackage{epsf}
\usepackage{graphics}
\usepackage{graphicx}

\begin{document}

\newcommand{\bc}{\begin{center}}
\newcommand{\ec}{\end{center}}
\newcommand{\be}{\begin{equation}}
\newcommand{\ee}{\end{equation}}
\newcommand{\bq}{\begin{eqnarray}}
\newcommand{\eq}{\end{eqnarray}}
\newcommand{\PLB}{{\it{Phys. Lett. {\bf{B}}}}}
\newcommand{\NPB}{{\it{Nucl. Phys. {\bf{B}}}}}
\newcommand{\PRD}{{\it{Phys. Rev. {\bf{D}}}}}
\newcommand{\AOP}{{\it{Ann. Phys. }}}
\newcommand{\PRL}{{\it{Phys. Rev. Lett. }}}
\newcommand{\MPL}{{\it{Mod. Phys. Lett. }}}

\title{Implicit Regularization and Renormalization of QCD}
\date{\today}

\author{Marcos Sampaio} \email []{msampaio@fisica.ufmg.br}
\author{ A. P. Ba\^eta Scarpelli} \email[]{scarp@fisica.ufmg.br}
\author{J. E. Ottoni}\email[]{jeottoni@fisica.ufmg.br}
\author{M. C. Nemes}\email []{carolina@fisica.ufmg.br}

\affiliation{Federal University of Minas Gerais -
Physics Department - ICEx \\
P.O. BOX 702, 30.161-970, Belo Horizonte MG - Brazil}

\begin{abstract}
\noindent

We apply the Implicit Regularization Technique (IR) in a non-abelian gauge
theory. We show that IR preserves gauge symmetry as encoded in relations
between the renormalization constants required by the Slavnov-Taylor identities
at the one loop level of QCD. Moreover, we show that the technique handles
divergencies in massive and massless QFT on  equal footing.

\end{abstract}
\pacs{11.15.-q, 11.15.Bt}

\maketitle

\section{Introduction and basics of Implicit Regularization}

Dimensional Regularization (DR) is the natural framework for computing Feynman
diagrams in gauge field theories. However
the regularization of dimension specific quantum field theories
such as chiral, topological and supersymmetric gauge  theories is known to
be a delicate matter in the context of DR. That is because the analytical
continuation on the space-time dimension of the Levi-Civita tensor is not
well-defined whereas supersymmetry is intrinsically defined on the physical
dimension of the underlying model.

Although some extensions of DR have been constructed (e.g. Dimensional
Reduction \cite{Siegel}) they are in general inconsistent in arbitrary loop
order and may give rise to spurious anomalies. Hence a judicious order by
order calculation in which the symmetry content of the model is
assured via constraint equations has to be performed. The drawbacks are clear:
in addition to turning the calculations cumbersome and tedious, we cannot rely
on such procedure to study anomalous (quantum mechanical) symmetry breaking.
This is particularly relevant in the supersymmetric
extensions of the standard model \cite{Jack}.

This motivates the search for a non-dimensional regularization/renormalization
scheme which, besides preserving the vital symmetries of the quantum field
theoretical model, is friendly from the calculational viewpoint.

Implicit Regularization (IR) is a momentum space setting to perform Feynman
diagram calculations in regularization independent fashion. The  Lagrangian of
the underlying quantum field theory is not modified: neither an explicit
regulator is introduced nor the dimensionality of the space time is moved away
from its physical dimension.  It has been
successfully applied  to various quantum field theoretical models including
those which make sense only in their physical dimension. For quantum
eletrodynamics, theories involving parity violating objects (Chern-Simons,
Chiral Schwinger Model), see \cite{PRD1}. For the study of anomalies and CPT
violation in an extended chiral version of quantum electrodynamics see
\cite{PRD2}. A comparison between IR, dimensional regularization, differential
renormalization and BPHZ  forest formula can be found in \cite{PRD3}, where it
is also calculated the beta function to one loop order in quantum
chromodynamics. In \cite{NLOOP} a model calculation using $\phi^3$ theory in $6$
dimensions illustrates how IR works when overlapping divergencies occur.  In \cite{JHEP}
it is  shown that IR  is manisfestly supersymmetric invariant.
This is illustrated  by renormalizing the massless Wess-Zumino model  and
calculating the beta function to three loop order. Application to a Gauged
Nambu-Jona Lasinio model can be found in \cite{CO}.

The main idea behind IR is very simple. The ultraviolet behaviour of the
amplitude is isolated as irreducible loop integrals (ILI's) which are
independent of the external momenta and need not be explicitly evaluated to
display the physical content of such amplitude. This can be achieved by
judiciously using the identity at the level of the integrand
\bq
&& \frac{1}{[(k+k_i)^2-m^2]} = \sum_{j=0}^{N} \frac{(-1)^j (k_i^2+ 2 k_i
\cdot k)^j}{(k^2-m^2)^{j+1}} \nonumber \\ &&+
\frac{(-1)^{N+1} (k_i^2 + 2 k_i \cdot k)^{N+1}}{(k^2 -m^2)^{N+1}
[(k+k_i)^2-m^2]} \, ,
\label{eqn:rr}
\eq
in order to eliminate the external momentum $k_i$ from the ILI,  $N$ being
chosen so that the last term is finite under integration over $k$.

You may assume, to be very strict, that a
regularization (say dimensional regularization) implicitly acts on the amplitude
in order to use (\ref{eqn:rr}) in the integrand. However once you have separated
the divergencies as irreducible loop integrals from the finite part of the
amplitude you need {\it{not}} compute the divergent integrals within IR. They
may be subtracted  and absorbed in the counterterms exactly as they stand.
The explicit computation of such ILI's is the origin of spurious symmetry
breaking which may contaminate the physics of the underlying model.
In \cite{PRD3} we have defined what is meant by a minimal subtraction scheme
within IR and compared with dimensional regularization and differential
renormalization.  In this process a natural renormalization group scale emerges
as it should. The generalization of this program to higher loop order is
straightforward: the overlapping divergences can be treated in a similar fashion
following a well-defined prescription which corresponds to the BPHZ forest
formula \cite{NLOOP}.

At the one loop level in the Minkowskian $4-$dimensional space time the ILI's
show up as
\be
\Theta^{(\infty)}_{\alpha_1, \alpha_1 ... \alpha_m} (\mu^2)= \int_k \frac{1,
k_{\alpha_1}, k_{\alpha_1}...k_{\alpha_m}}{(k^2-\mu^2)^n}
\ee
where $\int_k \equiv \int \frac{d^4 k}{(2 \pi)^4}$, $k$ being the internal
momentum, $\mu$ is an infrared regulator and $n=1, 2, ...$.  Typical higher
loops (logarithmically divergent) ILI's are, in $4$ dimensions
\be
\Theta^{n}_{\alpha_1, \alpha_1 ... \alpha_m} \equiv \int_k \frac{1,
k_{\alpha_1}, k_{\alpha_1}... k_{\alpha_m}}{(k^2-\mu^2)^p}  \ln^n \Big(
\frac{-\lambda^2}{k^2 - \mu^2}\Big) \, ,
\ee
where $4+m = 2p$ and $\lambda$ is an arbitrary dimensionful
non-vanishing constant originated in the previous order \cite{JHEP}.
Some comments are in order.   For massless models we may always introduce a
ficticious mass to regulate the propagators in the infrared limit without
sacrificing neither gauge symmetry \cite{PRD3} nor supersymmetry \cite{JHEP}.
We shall explicitly verify this in the context of QCD. Local arbitrary
counterterms will appear in IR as differences between irreducible loop integrals
of the same degree of divergence. Because we are not explicitly evaluating the
divergent integrals, such (finite) differences will have the status of free
parameters which should be adjusted by phenomenology or symmetry constraints.
Explicit regularizations will generally assign a (regularization dependent)
value to such differences which may lead to symmetry breaking.

In $3+1$ dimensions at the one loop level these arbitrary parameters look like:
\be
\Upsilon_{\mu \nu}^2 \equiv   g_{\mu \nu} I_{quad} (m^2) - 2 \Theta_{\mu
\nu}^{(2)}   = \alpha_1  g_{\mu \nu} \, ,
\label{CR4Q1}
\ee
\be
\Upsilon_{\mu \nu}^0 \equiv g_{\mu\nu} I_{log} (m^2) - 4 \Theta_{\mu \nu}^{(0)}
= \alpha_2 g_{\mu \nu}
 \label{CR4L1}
\ee
\be
\Upsilon_{\mu \nu \alpha \beta}^2 \equiv
g_{\{\mu\nu}g_{\alpha\beta \}} I_{quad} (m^2)  - 8 \Theta_{\mu \nu \alpha
\beta}^{(2)}   = \alpha_3 g_{\{\mu\nu}g_{\alpha\beta \}} \, ,
\label{CR4Q2}
\ee
\be
\Upsilon_{\mu \nu \alpha \beta}^0 \equiv
g_{\{\mu\nu}g_{\alpha\beta \}}  I_{log} (m^2)  - 24   \Theta_{\mu \nu \alpha
\beta}^{(0)}   =  \alpha_4 g_{\{\mu\nu}g_{\alpha\beta \}} \,
\label{CR4L2}
\ee
where
\begin{eqnarray}
I_{log} (m^2) &=& \int_k  \frac{1}{(k^2 - m^2)^2} \,\, ,\nonumber \\
I_{quad} (m^2) &=& \int_k  \frac{1}{(k^2 - m^2)} \,\, ,\nonumber \\
\Theta_{\mu \nu}^{(0)}(m^2)&=&\int_k  \frac{k_\mu k_\nu}{(k^2-m^2)^3} \,\, ,
\nonumber \\ \Theta^{(2)}_{\mu \nu} (m^2)&=& \int_k \frac{k_\mu
k_\nu}{(k^2-m^2)^2} \,\, ,\nonumber \\ \Theta_{\mu \nu \alpha \beta}^{(0)} (m^2)
&=& \int_k \frac{k_\mu k_\nu k_\alpha k_\beta}{(k^2-m^2)^4}
\,\, , \nonumber \\ \Theta_{\mu \nu \alpha \beta}^{(2)} (m^2) &=&
\int_k \frac{k_\mu k_\nu k_\alpha k_\beta}{(k^2-m^2)^3}
\, ,
\end{eqnarray}
$g_{\{\mu\nu}g_{\alpha\beta \}}$ stands for
$g_{\mu\nu}g_{\alpha\beta}+g_{\mu\alpha}g_{\nu\beta}+g_{\mu\beta}g_{\nu\alpha}$,
and the   $\alpha_i$'s are arbitrary, finite and regularization dependent.
Similar relations appear at higher loop order.

It is straightforward to see that  in dimensional regularization
(\ref{CR4Q1})-(\ref{CR4L2}) evaluate to zero. In \cite{PRD1} we have shown
that vector gauge symmetry is compatible with setting all the $\alpha_i$'s to
vanish. However we have shown that this is not the only solution. Such
feature somewhat explains why dimensional regularization is gauge invariant.

Whereas fixing the $\alpha_i$'s to zero from the start is more practical from
the calculational viewpoint care must be exercised
when  dealing with
dimension specific objects such as axial vertices and Levi-Civita tensors,
such arbitrary parameters should be fixed on physical grounds.  In \cite{PRD1}
we demonstrated that   (\ref{CR4Q1})-(\ref{CR4L2})  are connected  to momentum
routing invariance in a Feynman diagram. Should $\alpha_i$ vanish then the
amplitude is momentum routing invariant. The ideal arena to test this feature is
the study of  chiral anomalies: in perturbation theory
such anomalies manifest themselves  as a breaking of momentum routing
invariance \cite{JACKIWC}. In \cite{PRD2} we have studied the
Adler-Bardeen-Bell-Jackiw anomaly for arbitrary momentum routing and  seen
that IR consistently display the triangle chiral anomaly  in a scheme-free
fashion in a way that the anomaly appears in the vector and axial Ward
identities on equal footing. This is the best we can expect from a
regularization scheme \cite{JACKIW}. We have also seen that the mass spectrum of
the Chiral Schwinger model is undetermined by a arbitrary paremeter which in
perturbation theory corresponds to a finite arbitrary number expressed by the
difference between logarithmically divergent integrals in IR. This is what is
expected from a non-perturbative calculation \cite{PRD2}.

In other words we have seen that should such differences be set to zero
($\alpha_i \, 's = 0$) then the amplitude is momentum routing
invariant and (abelian) gauge invariant (although if they assume a non-vanishing
value it does not necessarily mean that gauge invariance is broken). When an
explicit form of the regulator is used
such differences are assigned a (regularization dependent) value. In
general we should keep any
arbitrariness which appears in perturbation theory until the final stage of the
calculations where physical conditions may fix its value. In this sense, IR
is especially taylored to implement this idea especially when quantum symmetry
breakings may occur.

In \cite{PRD3} we have verified that constraining $\alpha_i$ to zero also
ensures the transversality of the vacuum polarization tensor of QCD.
The next stringent test for  establishing the generality of IR is to extend
these ideas to  nonabelian gauge field theories.

The purpose of this paper is threefold:

1) To check whether a constrained version of IR (CIR) generalizes to a
nonabelian gauge theory (QCD) and show that gauge symmetry is preserved as
expressed by the Slavnov-Taylor identities between the renormalization
constants and calculate all the renormalization group constants to one loop order;

 2) To define a renormalization group scale in the following way:
just as in the  case of the photon propagator, we introduce a ficticious mass
($\mu$) for the gluon which will appear in both the finite and the
(logarithmically) divergent pieces of the amplitude. For the divergent piece it
will show up as  the ILI
$$
I_{log} (\mu^2) = \int \frac{d^4 k}{(2 \pi)^4} \frac{1}{(k^2 - \mu^2)^2} .
$$
We eliminate the infrared mass regulator from the definition of the counterterm
by using the identity (see appendix A)
\be
I_{log} (\mu^2) = I_{log} (\lambda^2) + b\,  \ln \Big( \frac{\lambda^2}{\mu^2}
\Big) \,  ,
\label{eqn:Ilog}
\ee
\be
b \equiv i /(4 \pi)^2
\ee
and $\lambda$ is a nonvanishing parameter which
parametrizes the arbitrariness in separating the divergent from the finite
content of the amplitude and will play the role of a renormalization group scale
in IR.  As a byproduct we realize what is meant by a minimal subtraction, mass
independent scheme in IR, namely subtracting $I_{log} (\lambda^2)$ . The
infrared divergent term expressed by $\ln \mu^2$ will exactly cancel the
infrared cutoff dependence in the {\it{finite}} part of the amplitude, as it
should, for all infrared safe theories.

3) to see that unlike Dimensional Regularization, the tadpole graphs of Yang-Mills
fields play a crucial role for maintaining manifest gauge invariance through
cancellations of quadratic divergences which appear at one loop order.

\section{1 Loop QCD in Implicit Regularization}

The QCD Lagrangian reads

\begin{eqnarray}
{\cal{L}}_0 &=& \frac{1}{4}(F_{ 0 \mu \nu }^{a})^{2}-\frac{1}{2\alpha }
(\partial ^{\mu }A_{0\mu }^{a})^{2} \nonumber \\ &+&
\bar{\psi}_{0}^{i} (i \gamma_\mu D_\mu^{ij} - m_0 \delta^{ij}) \psi_0^j  + \\
&+& i (\partial^\mu \bar{c}_0^a)D_\mu^{ab}c_0^b
\label{eqn:lag}
\end{eqnarray}
with $D_\mu^{ab}$ and $D_\mu^{ij}$ referring to the adjoint and fundamental
representation of the colour group  $SU(3)$. Also
$$
F_{0\mu \nu }^{a}=\partial _{\mu }A_{0\nu
}^{a}-\partial _{\nu }A_{0\mu }^{a}+g_{0}f^{abc}A_{0\mu }^{b}A_{0\nu }^{c},
$$
$$
D_\mu = (\partial _{\mu }-ig_{0 1}A_{0\mu
}^{a}t_{r}^{a}) \, ,
$$
$\alpha$  is the gauge-fixing parameter and
$A_{0 \mu}$ are the gauge fields coupled to $n_f$ Dirac fermions $\psi_0$ and to
the ghost fields $c_0$. The index  ``0" stands for bare quantities.  The group
theoretical factors which will appear in the amplitudes are defined through
the relations $tr (t^a_r t^b_r)= C(r) \delta^{ab}$, $t^a_r  t^a_r = C_2(r)
{\bf{\hat{1}}}$, $f^{acd} f^{bcd} = C_2 (G) \delta^{ab}$.  Because the
interaction terms in the Lagrangian above are interrelated by BRS symmetry, only
one coupling constant is left independent. Consequently the renormalization
constants will be constrained by generalized Ward-Takahashi identities (
Slavnov-Taylor identities).

We define renormalized fields and couplings through the renormalization
constants as below
\be
A_{0 \mu}^a = Z_3^{1/2} A_\mu^a \,\,\, , \,\,\, c_0^a = \tilde{Z}_3^{1/2} c^a
\,\,\, , \,\,\, \psi_0 = Z_2^{1/2} \psi \,\,\, ,
\nonumber
\ee
\be
g_0 = Z_g g \,\,\, , \,\,\, m_0 = Z_m  m \, .
\label{eqn:bare}
\ee
Therefore we may define ${\cal{L}}_0 = {\cal{L}} + {\cal{L}}_{ct}$ where
${\cal{L}}$ is precisely equal to  ${\cal{L}}_0$  except that it is written in
terms of the renormalized variables whereas    ${\cal{L}}_{ct}$ is the
counterterm Lagrangian which reads
\bq
{\cal{L}}_{ct} &=& (Z_3-1)\frac{1}{2} A_a^\mu \delta^{ab} (g_{\mu \nu}
\partial^2 - \partial_\mu \partial_\nu)A_b^\nu  + \nonumber \\
&+& (\tilde{Z}_3-1)\bar{c}^a\delta_{ab}(-i \partial^2) c^b  \nonumber \\ &+&
(Z_2 - 1)\bar{\psi}^i (i \gamma^\mu \partial_\mu - m) \psi^i  \nonumber \\
&-& (Z_2 Z_m -1) m \bar{\psi}^i \psi^i \nonumber \\  &-& (Z_1-1) \frac{1}{2} g
f^{abc} (\partial_\mu A^a_\nu - \partial_\nu A^a_\mu) A_b^\mu A_c^\nu
\nonumber \\ &-& (Z_4-1) \frac{1}{4} g^2 f^{abe}f^{cde} A^a_\mu A^b_\nu A_
c^\mu A_ d^\nu \nonumber \\ &-& (\tilde{Z}_1 -1) i g f^{abc} (\partial^\mu
\bar{c}^a) c^b A^c_\mu \nonumber \\ &+& (Z_{1F} -1)g\bar{\psi}^i t^a_{ij}
\gamma^\mu \psi^j A^a_\mu  \, ,
\eq
where we have defined
$$
Z_1 \equiv Z_g Z_3^{3/2} \,\,\, , \,\,\, Z_4 \equiv Z_g^2 Z_3^2 \,\,\, ,
$$
$$
\tilde{Z}_1 \equiv Z_g \tilde{Z}_3 Z_3^{1/2} \,\,\, , \,\,\, Z_{1F} \equiv Z_g
Z_2 Z_3^{1/2} \, .
$$
The equality of $Z_g$ for all the couplings leads to the Slavnov-Taylor
identities:
\be
\frac{Z_1}{Z_3}=\frac{\tilde{Z}_1}{\tilde{Z}_3}=\frac{Z_{1F}}{Z_2}=\frac{Z_4}{Z_
1} \,  .
\label{eqn:sti}
\ee
The Feynman rules for QCD can be found in
any textbook. We follow T. Muta \cite{Muta} and work in the
Feynman gauge where $\alpha = 1$. To one loop order, the relevant amplitudes are
represented by well-known diagrams which we shall call as
\bq
\Pi^{ab}_{\mu \nu}  &\rightarrow&  {\mbox{gluon self-energy}} \nonumber \\
\Sigma &\rightarrow& {\mbox{quark self-energy}} \nonumber \\
\Sigma^{ab}_{ghost} &\rightarrow&   {\mbox{ghost self-energy}} \nonumber \\
\Lambda^a_{\mu} &\rightarrow& {\mbox{quark-gluon vertex}} \nonumber \\
\Lambda^{abc}_{\mu \nu \lambda} &\rightarrow& {\mbox{three-gluon vertex}}
\nonumber \\
\Lambda_\mu^{abc} &\rightarrow& {\mbox{ghost-gluon vertex}} \nonumber \\
\Lambda^{abcd}_{\mu \nu \alpha \beta} &\rightarrow& {\mbox{four-gluon vertex}}
\, .
\eq

We start with the gluon self-energy which is composed of  four
contributions as depicted in fig. (\ref{fig1}).
\begin{equation}
\Pi _{\mu \nu }^{ab}=\Pi _{\mu \nu }^{ab}(1)+\Pi _{\mu \nu }^{ab}(2)+\Pi
_{\mu \nu }^{ab}(3)+\Pi _{\mu \nu }^{ab}(4),  \label{pimini}
\end{equation}
where $\Pi _{\mu \nu }^{ab}(1)$, $\Pi _{\mu \nu }^{ab}(2)$, $\Pi _{\mu \nu
}^{ab}(3)$ and $\Pi _{\mu \nu }^{ab}(4)$ represent the quark loop, the
gluon loop, the gluon tadpole and the ghost loop respectively.
It is purely transversal as
required by the Slavnov-Taylor identities and thus it does not admit a mass
term and there should be no mass renormalization. Hence the quadratic
divergences which appear in    $\Pi _{\mu \nu }^{ab}$ should cancel out.

\begin{eqnarray}
\Pi^{ab}_{\mu \nu}(1) &=& -g^2C_2(G) \delta^{ab}3\int_k \frac{g_{\mu \nu}}
{k^2-\mu^2}\nonumber \\
&=&    -3 g^2 g_{\mu \nu} C_2(G) \delta^{ab} I_{quad}(\mu^2)
\end{eqnarray}
where $\mu$ is an infrared cutoff (mass regulator) which should be set to zero
in the end.

The gluon loop amplitude reads
\begin{equation}
\Pi^{ab}_{\mu \nu}(2)= \frac{1}{2} \int_k  g^2 f^{acd} f^{bcd} N_{\mu \nu}
\frac{-i}{k^2-\mu^2}\frac{-i}{(k+p)^2-\mu^2} \,\, ,
\label{eqn:Pmn2}
\end{equation}
where
\begin{eqnarray}
N^{\mu \nu}&=&[g^{\mu \rho}(p-k)^{\sigma}+g^{\rho \sigma}(2k+p)^{\mu} +
g^{\sigma \mu}(-k-2p)^{\rho}]  \nonumber \\
&\times&
[\delta^{\nu}_\rho (k-p)_{\sigma}+g_{\rho \sigma}(-2k-p)^{\nu}+
\delta^{\nu}_{\sigma}(k+2p)_\rho]  \nonumber \\
&=& 2p_\mu p_\nu -5(p_\mu k_\nu +p_\nu k_\mu) -10 k_\mu k_\nu \nonumber \\
&-& g_{\mu \nu}[(p-k)^2+ (k+2p)^2] \, .
\end{eqnarray}
Using that
\begin{equation}
(p-k)^2+ (k+2p)^2=(k+p)^2+k^2+4p^2 ,
\end{equation}
(\ref{eqn:Pmn2}) may be cast as
\begin{eqnarray}
\Pi^{ab}_{\mu \nu}(2)&=& -\frac 12 g^2 C_2(G) \delta^{ab} [(2p_\mu p_\nu -
4p^2 g_{\mu \nu})J (p^2,\mu^2)  \nonumber \\
&-& g_{\mu \nu}(2 I_{quad}(\mu^2) + p^\alpha p^\beta
\Upsilon_{\alpha \beta}^0)   \nonumber \\
&-& 10(\, p_\nu J_\mu (p^2,\mu^2) + J_{\mu \nu}(p^2,\mu^2) \, )] \,  .
\end{eqnarray}
As for  the ghost loop, we have
\begin{eqnarray}
\Pi^{ab}_{\mu \nu}(3)&=& - g^2f^{dac}f^{cbd}  \int_k \frac{i^2}{k^2-\mu^2}
\frac { (p+k)_\mu k_\nu}{[(k+p)^2-\mu^2]}\nonumber \\
&=& - g^2 \delta^{ab} C_2(G) (p_\nu J_\mu  (p^2,\mu^2) \nonumber \\ &+& J_{\mu
\nu} (p^2,\mu^2) ) \,\, , \end{eqnarray}
in which $J_{\mu \nu}$, $J_\mu$ and $J$ are defined as
\begin{eqnarray}
&& J_{\mu \nu} (p^2, \mu^2) = \Theta^{(2)}_{\mu \nu}-
 p^2 \Theta^{(0)}_{\mu \nu}
+ 4 p^\alpha p^\beta \Theta^{(0)} _{\mu \nu \alpha \beta}    \\
&& + b \Bigg\{ \frac {p_\mu p_\nu}{3} \Big[\frac 16 -\frac
1{p^2}(p^2-\mu^2)Z (p^2,\mu^2) \Big]\\ &&- \frac{p^2 g_{\mu \nu}}{6}\Big[\frac
13 +\frac 1{2p^2}(-p^2+4 \mu^2)Z (p^2,\mu^2)\Big] \Bigg\} \, ,\\
&& J_\mu (p^2,\mu^2) = -2p^\alpha
\Theta_{\alpha \mu}^{(0)} + \frac b2 \, p_\mu  \, Z (p^2,\mu^2) \, ,\\ && J
= I_{log}(\mu^2) -b Z (p^2, \mu^2),
\eq
with
\be
Z (p^2, \mu^2) =  \int_0^1 dz \,\, \ln \Bigg( \frac{p^2 z
(1-z)-\mu^2}{-\mu^2} \Bigg),
\ee
which, for $\mu^2 \to 0$, is given by
\be
\ln {\left(-\frac{p^2}{e^2 \mu^2}\right)} \, .
\ee

Collecting  all the results so far enables us to write
\bq
&&\sum_{i=1}^3 \Pi^{ab}_{\mu \nu}(i) = g^2 C_2(G) \delta^{ab} \Bigg\{
\Big(p^2 g_{\mu \nu} -p_\mu p_\nu \Big) \times \nonumber \\
&&\times \Bigg[- \frac {2b}{9} + \frac 53 \Bigg( I_{log}(\mu^2)- b
\ln \Big( -\frac{p^2}{ e^2 \mu^2} \Big) \Bigg) \Bigg] \nonumber \\
&&+ \Upsilon^2_{\mu \nu} + p^2 \Upsilon^{(0)}_{\mu \nu} +
p^\alpha p^\beta \Upsilon^{(0)}_{\mu \nu \alpha \beta} + \nonumber \\
&& + p^\alpha p_\mu \Upsilon^{(0)}_{\nu \alpha} + p^\beta
p_\nu \Upsilon^{(0)}_{\mu \beta} + p^\alpha p^\beta  g_{\mu
\nu} \Upsilon^{(0)}_{\alpha \beta} \Bigg\} \, ,
\label{eqn:Pmn123}
\eq
in which the $\Upsilon$'s are the arbitrary constants defined in  the relations
(\ref{CR4Q1}) to (\ref{CR4L2}). Moreover in writing (\ref{eqn:Pmn123}) we have
absorbed  some constant factors in the $\Upsilon$'s.

The fermion loop contribution to the gluon self energy is identical to the
vacuum  polarization tensor of $QED$ except for the colour and number of
fermions ($n_f$) factors. It has been computed within IR elsewhere \cite{PRD2}.
Without loss of generality we write the result in the limit of massless 
fermions to yield
\begin{eqnarray}
&&\Pi_{\mu \nu}^{ab}(4) = \frac{4}{3} g^2  C(r) n_f
\delta^{ab}   \Bigg\{       \Big(p_\mu p_\nu-p^2 g_{\mu \nu} \Big)  \times
\nonumber \\ && \times
\Bigg[ I_{log} (\mu^2) - b  \Bigg( \ln \Big(-\frac{p^2}{e^2 \mu^2} \Big)
+ \frac{1}{3} \Bigg) \Bigg]  \nonumber \\
&& + \Upsilon^2_{\mu \nu} + p^2 \Upsilon^{(0)}_{\mu \nu} +
p^\alpha p^\beta \Upsilon^{(0)}_{\mu \nu \alpha \beta} + \nonumber \\
&& + p^\alpha p_\mu \Upsilon^{(0)}_{\nu \alpha} + p^\beta
p_\nu \Upsilon^{(0)}_{\mu \beta} + p^\alpha p^\beta  g_{\mu
\nu} \Upsilon^{(0)}_{\alpha \beta}  \Bigg\} \, .
\end{eqnarray}

Some comments are in order. Firstly notice that the quadratic divergences
expressed by $g _{\mu \nu}I_{quad} (\mu^2)$ and $\Theta_{\mu \nu}^2$ that appear
in the gluon tadpole, the gluon loop and the ghost loop amplitudes combine to
make up $\Upsilon^2_{\mu \nu}\equiv \alpha_1 g_{\mu \nu}$.  Gauge invariance
tells us that we ought to set $\alpha_1=0$ as well as all the others
$\alpha_i$'s as defined in (\ref{CR4Q1})-(\ref{CR4L2}). Hence tadpole graphs
of gauge fields play an essential role in maintaining gauge invariance within
our framework. DR automatically sets quadratic divergences to zero in the limit
where $\mu \rightarrow 0$. Here this is not necessary in order to ensure the
transverse form of the gluon self-energy as required by gauge invariance. As we
shall see, setting $\lambda_i$'s to zero in (\ref{CR4Q1}) to (\ref{CR4L2})
automatically preserves (vector) gauge
invariance through the Slavnov-Taylor identities. This is in consonance with the
idea that ultimately one should let arbitrary parameters to be fixed on physical
grounds. In the present case, gauge invariance does this job. However they were
shown to play a crucial role in describing correctly chiral field theories
in which the Dirac algebra involving $\gamma_5$ matrices prevents the use of
naive DR. In recent work \cite{Andre}, such free parameters have been taken into
account in a renormalized version of a $SU(3)$ Nambu and Jona-Lasinio like
model.   The relevant observables have been calculated in excelent agreement
with experiment including a simultaneous and satisfactory fit for both $f_\pi$
and $f_\kappa$. This is an interesting feature because it enables IR to be
applicable to study the dynamics of effective field theories (for instance the
derivation of the gap equation in the gauged Nambu and Jona-Lasinio model
\cite{NJL} in a gauge invariant fashion). Moreover the leading quadratic terms
also play a crucial role in the evaluation of the hadronic matrix elements of
four quark operators in the kaon decays $K \rightarrow \pi \pi$  as well as in
providing a consistent prediction on the direct CP violating parameter
$\epsilon'/\epsilon$ in kaon decays \cite{Kaon}. IR may be applied to all these
scenarios. Operationally it is convenient because one has a gauge invariant
momentum space framework.

Note that the algebraic procedure that we have used to define
a mass independent, minimal, renormalization scheme naturally introduces an
arbitrary scale $\lambda$.  As we shall see $\lambda$ plays the role of a
renormalization group scale.

In order to define genuine renormalization
constants which display the ultraviolet scaling behaviour of the model we use
the identity (\ref{eqn:Ilog}) and note that the (infrared) divergence
parametrized by $\ln \mu^2$ as $\mu \rightarrow 0$ cancels out against an
identical term coming from the UV finite part whilst an arbitrary nonvanishing
parameter $\lambda$ appears.  Altogether $\Pi _{\mu \nu }^{ab} =
\sum_{i=1}^4 \Pi_{\mu \nu}^{ab}(i)$ reads
\begin{eqnarray}
&& \Pi _{\mu \nu }^{ab}(p^2, \lambda^2)) = -\frac{b}{9} g^{2}(p^{2}g_{\mu \nu
}-p_{\mu }p_{\nu })\delta ^{ab}\times   \notag \\ &&\times \Bigg\{
i \Big[ \frac{5}{3}C_{2}(G)-\frac{4}{3}n_{f}C(r) \Big] I_{\log }(\lambda ^{2})
\nonumber \\ && +\Big( 15C_{2}(r)-6n_{f} \Big) \ln \Big(\frac{\lambda
^{2}}{p^{2}}\Big)-2C_{2} (r)+2 n_{f} \Bigg\}  \nonumber \\
&& +(Z_3 -1) \delta^{ab} (p^{2}g_{\mu \nu}-p_{\mu }p_{\nu })\, .
\label{eqn:Pmnf}
\end{eqnarray}

We define the counterterm for the amplitude (\ref{eqn:Pmnf}) by minimally
subtracting (in the IR sense) the ILI expressed by $I_{log} (\lambda^2)$
to define
\begin{equation}
Z_{3}=1-i\left[ \frac{5}{3}C_{2}(G)-\frac{4}{3}n_{f}C(r)\right] I_{\log
}(\lambda ^{2})g^{2}+O(g^{3}).
\label{z3}
\end{equation}
Note that the algebraic procedure that we have used to define
a mass independent, minimal, renormalization scheme naturally introduces an
arbitrary scale $\lambda$.  As we shall see $\lambda$ plays the role of a
renormalization group scale.
In \cite{PRD3} we compare renormalization schemes in IR, DR and differential
renormalization (see also \cite{Dunne}).

\begin{figure}[h]
\centerline{\hbox{
   \epsfxsize=8in
   \epsfysize=8in
   \epsffile{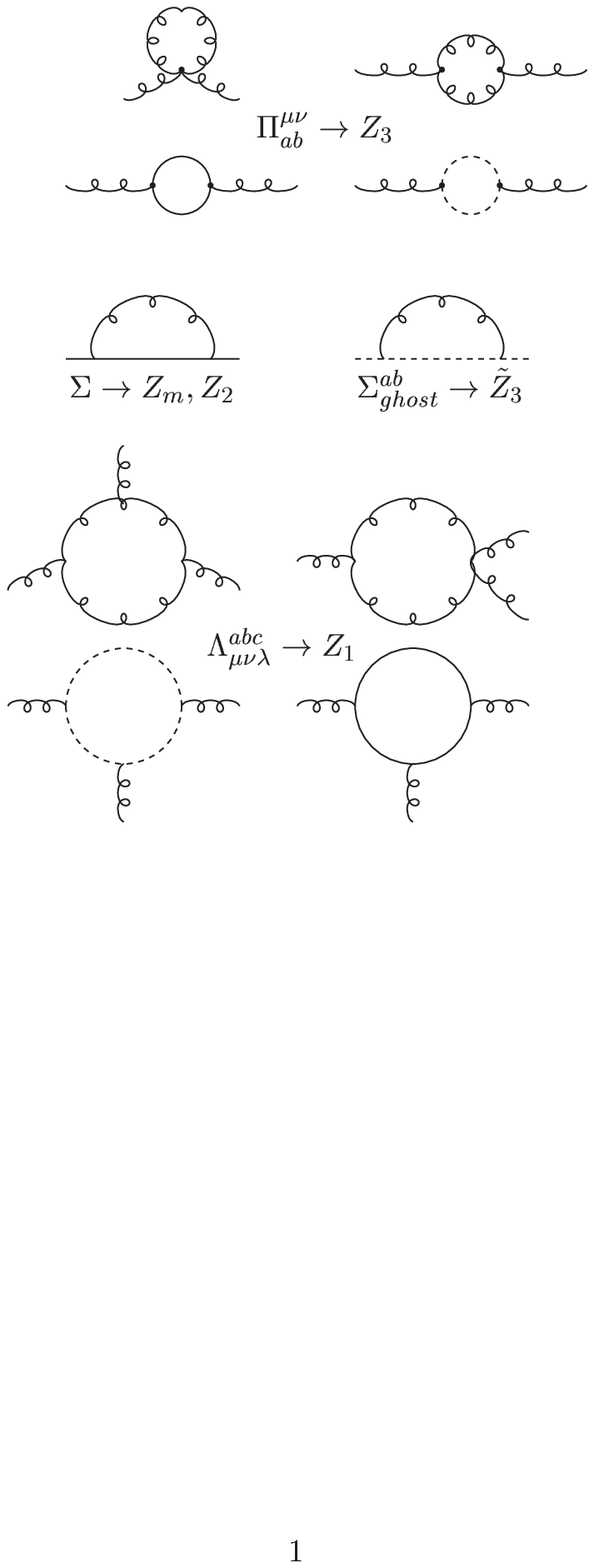}
     }
  }
  \vskip-7.5cm
  \caption{1 loop diagrams of QCD
          }
  \label{fig1}

\end{figure}

 \begin{figure}[h]

   \centerline{\hbox{
   \epsfxsize=8in
   \epsfysize=8in
   \epsffile{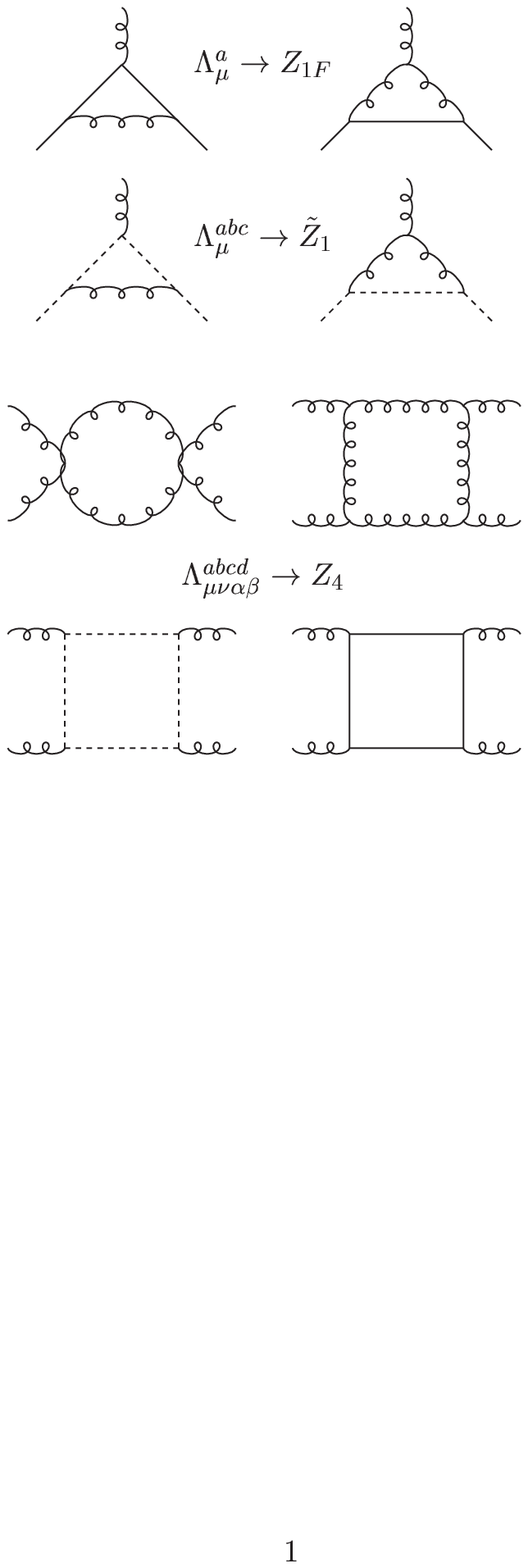}
     }
  }
  \vskip-7.5cm
  \caption{1 loop diagrams  of QCD
          }
\label{fig2}

\end{figure}

The quark self energy $\Sigma (p)$ is similar
to the electron self energy apart from a group theoretical factor. $\Sigma (p)$
has been calculated in \cite{PRD3} within IR. It reads
\bq
\Sigma (p) &=& - i g^{2}C_{2}(r) ( p \hspace{-2.3mm}/ - 4 m) I_{\log }(m^2
)+ g^2 g^{\mu \nu} \Upsilon_{\mu \nu}^{(0)}  p \hspace{-2.3mm}/ \nonumber \\ &+& (Z_2 -1) p
\hspace{-2.3mm}/  - (Z_2 Z_m - 1)m + \widetilde{\Sigma} (p, m) \,
, \eq
where the tilde means that the quantity is finite. We shall use such notation
from now on. Indeed  $\widetilde{\Sigma} (p, m)$ is both ultraviolet and
infrared finite. In order to define the corresponding renormalization constants
we consistently make use of (\ref{eqn:Ilog})  in order to pursuit a mass
independent scheme as well as introducing the arbitrary constant $\lambda^2$.
It is noteworthy that the $b \ln \mu^2$ piece coming from (\ref{eqn:Ilog})
cancels exactly the infrared divergence in the ultraviolet finite portion of the
amplitude, as it should.

Henceforth we shall systematically define the renormalization constants in a
mass independent fashion (which defines the ``minimal" scheme in IR) as well as
set $\alpha_i$'s ($\Upsilon$'s) to zero. We define this procedure   as
constrained IR (CIR).  We shall no longer write the $\Upsilon$'s explicitly in
the remaining amplitudes for the sake of brevity.

Therefore the fermion mass and field renormalization
constants can be  cast as
\bq
Z_{m} &=&1 + 3 i C_{2}(r) I_{\log }(\lambda ^{2}) g^{2}+O(g^{3}) \, ,
\nonumber \\
Z_{2} &=& 1+ i C_{2}(r) I_{\log }(\lambda ^{2})g^{2}+O(g^{3}) \, .
\label{eqn:ZmZ2}
\eq

After the appropriate color index contractions, the one loop correction for the
ghost propagator simplifies to
\bq
\Sigma _{ghost}^{ab}(p^2) &=& g^{2}g_{\mu \nu }C_{2}(G)\delta
^{ab}\int_{k}\frac{ (p-k)^{\mu }p^{\nu
}}{[(p-k)^{2}-\mu^{2}](k^{2}-\mu^{2})}\nonumber \\
&+& ({\tilde{Z_3}}-1) \delta^{ab}p^2 \, ,
\eq
where $\mu^2$ is an infrared mass regulator for both the ghost
and gluon propagators. After some
straightforward algebra  we have
\bq
\Sigma _{ghost}^{ab}(p^2, \lambda^2) &=& \delta ^{ab} p^2 \Big(
\frac{i g^2}{2} C_2(G) I_{\log }(\lambda ^{2}) \nonumber \\ &+&
{\tilde{Z_3}} -1 \Big) + \widetilde{\Sigma}^{ab} _{ghost} (p^2, \lambda^2)  \, ,
\eq
from which we define the renormalization constant
\begin{equation}
\tilde{Z}_{3}=1- \frac{i}{2}C_{2}(G)I_{\log }(\lambda ^{2})g^{2}+O(g^{3}).
\label{eqn:Z3t}
\end{equation}

For the one loop quark-gluon vertex $\Lambda^{a}_{\mu }$ shown in fig.
(\ref{fig2}) we have two contributions:  the QED-like electron-photon vertex
diagram $\Lambda^{a}_{\mu }(p, q) (1)$ and the one involving the three-gluon
vertex $\Lambda^{a}_{\mu }(p, q) (2)$. The former
differs from the QED electron-photon vertex  by a group theoretical factor
\be
\Lambda^{a}_{\mu} (p, q) (1)  = (t^d t^a t^d) \Lambda_{\mu
}^{QED} (p, q) \, .
\ee
We have also computed $\Lambda_{\mu}^{QED} (p,q)$ within IR in \cite{PRD3} so
here we only quote the result:
\be
i \Lambda_{\mu}^{QED} (p, q) = \gamma^\mu [ \alpha_2  +  I_{log} (m^2) ] +
\tilde{\Lambda}^\mu (p, q) \, ,
\ee
where $m$ is the mass of the fermion, $\tilde{\Lambda}^\mu (p, q) $ is finite
and $\alpha_2$  is arbitrary and shall be set to zero within
constrained IR.  Using that $t^d t^a t^d = [ C_2(r) -1/2 \, C_2 (G) ] t^a$
enables us to write
\bq
&& \Lambda^{a}_{\mu}(p, q, \lambda^2) (1) = - i g^3 t^a  \gamma_\mu \Big( C_2
(r) - \frac{1}{2} C_2 (G) \Big) \times  \nonumber \\ && \times \Big(
I_{log} (\lambda^2) + \tilde{\Lambda}^{a} (p, q,\lambda^2) (1) \Big) \, .
\eq
As for  $\Lambda^{a}_{\mu} (p, q) (2)$, the Feynman rules give
\be
i \Lambda ^{a}_{\mu} (p, q) (2) = g^3 f^{abc} t^b t^c \int_k
\frac{{\cal_{N}}_\mu}{{\cal{D}}}
\ee
with
\bq
{\cal{N}}_\rho &=& \gamma^\mu (k\hspace{-2.3mm}/ + m) \gamma^\nu \Big(
(2 p - q - k)_\nu g_{\mu \rho} \nonumber \\ &+& (2 k - p - q)_\rho g_{\mu \nu}
+ (2 q - p - k)_\mu g_{\nu \rho} \Big) \, , \nonumber \\
{\cal{D}} &=& (k^2 -m^2)[(k-p)^2 - \mu^2] [(k-q)^2 - \mu^2] \, .
\eq
where again $\mu$  is a mass regulator for the gluon propagator. We proceed as
before. We remove the external momentum dependence from the ILI by applying
the identity (\ref{eqn:rr}) in the propagators which contain the momenta $p$ and
$q$ above. Then we  isolate a genuine (ultraviolet divergent only) contribution
for the counterterm with the help of identity   (\ref{eqn:Ilog}) to get, in CIR,
\bq
\Lambda ^{a}_{\mu} (p, q, \lambda^2) (2) &=& -i \frac{3}{2} g^3 C_2 (G) t^a
\Big( \gamma_\mu I_{log} (\lambda^2) \nonumber \\ &+& \tilde{\Lambda}^{a}_{\mu}
(p, q, \lambda^2) (2)  \Big)\, .
\eq
Finally we define the renormalization constant $Z_{1F}$ to one loop order by
adding up the two contributions for the quark-gluon vertex:
\bq
\Lambda ^{a}_{\mu} (p, q, \lambda^2)  &=&  g \gamma_\mu t^a \Big(
 - i g^2 (C_2(G)+C_2(r)) I_{log}(\lambda^2) + \nonumber \\
 &+& Z_{1F} - 1 + \widetilde{\Lambda} ^{a} (p, q, \lambda^2) \Big)
\eq
which gives
\be
Z_{1F} = 1 + i g^2 \Big( C_2 (G) + C_2 (r) \Big) I_{log} (\lambda^2)     \, .
\ee

To calculate the renormalization constant $\tilde{Z}_1$ we work with
the ghost-gluon vertex. It receives two contributions (fig. \ref{fig2}): the
ghost-ghost-gluon loop ${\Lambda}^{abc}_{1 \mu}$ and the ghost-gluon-gluon
loop  ${\Lambda}^{abc}_{2 \mu}$. Let  $p_1$, $p_2$ and $p_1 + p_2 \equiv
q$ be the external momenta.  Thus
\be
{\Lambda}^{abc}_{\mu} = \tilde{\Lambda}^{abc}_{1 \mu} +
{\Lambda}^{abc}_{2 \mu} + (-i)g f^{abc}q_\mu ({\tilde{Z_1}}-1)  \, .
\ee
It is straightforward to use the Feynman rules with the help of the identity
$f^{afe}f^{bep}f^{cpf} = N/2 f^{abc}$ for $SU(N)$ to get
\bq
&& {\Lambda}^{abc}_{1 \mu} (p_1, p_2) = - \frac{g^3}{2} C_2(G) f^{abc}
q^\alpha \nonumber \\
&& \int_k  \frac{k_\alpha
k_\mu}{(k^2-\mu^2)[(k-p_1)^2-\mu^2][(k-q)^2-\mu^2]} \, .
\eq
We proceed according to the rules of CIR, as we have done before, to arrive at
\be
{\Lambda}^{abc}_{1 \mu} (p_1, p_2) = \frac{g^3}{8} C_2(G)  f^{abc} q_\mu
I_{log} (\lambda^2) +     \tilde{{\Lambda}}^{abc}_{1 \mu} (p_1, p_2)\, .
\ee
Similarly we have for the other contribution
\be
{\Lambda}^{abc}_{2 \mu} (p_1, p_2) = \frac{3 g^3}{8} C_2(G)  f^{abc} q_\mu
I_{log} (\lambda^2) +     \tilde{{\Lambda}}^{abc}_{2 \mu} (p_1, p_2)\, ,
\ee
and hence
\bq
{\Lambda}^{abc}_{\mu} (p_1, p_2) &=& -i g  f^{abc} q_\mu \Big(
-i \frac{g^2}{2} C_2(G) I_{log} (\lambda^2) +   \tilde{Z}_1 - 1 \Big) \nonumber
\\ &+& \tilde{{\Lambda}}^{abc}_{\mu} (p_1, p_2)\, .
\eq
Finally we define the renormalization constant in a minimal fashion within IR as
\be
\tilde{Z}_1 = 1 + \frac{i}{2} g^2 C_2(G) I_{log} (\lambda^2)\, .
\ee

The class of one loop three-gluon vertex graphs from which we shall define
$Z_ 1$ are shown in fig. \ref{fig1}. They have been explicitly computed  in
\cite{Celmaster}, \cite{Pascual} within DR. The calculation is straightforward
yet tedious. We have proceeded according to the rules of CIR as before in order
to isolate the ultraviolet divergence as a term proportional to $I_{log}
(\lambda^2)$ after making use of (\ref{eqn:Ilog}). The infrared divergent piece
proportional to $\ln (\lambda^2/\mu^2)$ as $\mu \rightarrow 0$ cancels out with
an alike term stemming from the ultraviolet finite piece of the amplitude as we
generically discuss in the appendix B.  For the sake of brevity
we shall present only the result here. Let $p$ and $q$ be the external momenta.
Then
\bq
\Lambda^{abc}_{\mu \nu \lambda}(p,q) &=& -i g f^{abc}V_{\mu \nu
\lambda}(p,q,p+q) \times \nonumber \\ &\times& \Bigg(- i g^2 \Big( -\frac{2}{3}
C_2(G) + \frac{4}{3} C(r) n_f \Big) I_{log} (\lambda^2) \nonumber \\ &+&  Z_1 -
1 \Bigg) + \widetilde{\Lambda}^{abc}_{\mu \nu \lambda}(p,q)      \, ,
\eq
$V_{\mu \nu \lambda}(p,q,p+q) = (p-q)_\lambda g_{\mu \nu} -p_{\mu} g_{\nu
\lambda} + q_{\nu} g_{\mu \lambda} $,  from which we define
\be
Z_1 = 1 + i g^2 \Bigg(- \frac{2}{3} C_2(G) + \frac{4}{3} C(r) n_f
\Bigg)I_{log}(\lambda^2) \, .
\ee

Last but not least we have to compute the four gluon vertices  depicted in
fig. \ref{fig2} along with all their permutations. This long
calculation has been performed with great detail by Pascual and Tarrach in
\cite{Pascual} in the Weinberg's scheme as well as by Papavassiliou in
\cite{Papavassiliou} using the S-matrix pinch technique . The
corresponding result within CIR reads:
\bq && \Lambda^{abcd}_{\alpha \beta \mu \nu} (p_1,p_2,p_3,p_4) = -g^2
W^{abcd}_{\alpha \beta \mu \nu} \times \nonumber \\ && \times \Bigg(
- \frac{i g^2}{3} \Big( C_2(G) + 4 C(r) n_f \Big) I_{log}(\lambda^2)
+ Z_4 -1 \Bigg) \nonumber \\ && + \, \widetilde{\Lambda}^{abcd}_{\alpha \beta
\mu \nu} (p_1,p_2,p_3,p_4) \, ,
\eq
where we have used the same notation as T. Muta \cite{Muta},  namely
\bq
W^{a_1 a_2 a_3 a_4}_{\mu_1 \mu_2 \mu_3 \mu_4} &=&
(f^{13,24}-f^{14,32})g_{\mu_1 \mu_2} g_{\mu_3 \mu_4} \nonumber \\
&+& (f^{12,34}-f^{14,23})g_{\mu_1 \mu_3} g_{\mu_2 \mu_4} \nonumber \\
&+& (f^{13,42}-f^{12,34})g_{\mu_1 \mu_4} g_{\mu_3 \mu_2} \, ,
\eq
with $f^{ij,lm}=f^{a_i a_j a}f^{a a_l a_m} $.  Therefore
\be
Z_4 = 1 + i g^2 \Bigg( \frac{1}{3}C_2(G) + \frac{4}{3} C(r) n_f \Bigg)
I_{log}(\lambda^2) \, .
\ee

\section{Slavnov-Taylor Identities and Renormalization Group Functions}

It is a simple task to verify that CIR explicitly preserves the Slavnov-Taylor
identities expressed by (\ref{eqn:sti}):
\be
\frac{Z_1}{Z_3}=\frac{\tilde{Z}_1}{\tilde{Z}_3}=\frac{Z_{1F}}{Z_2}=\frac{Z_4}{Z_
1} = 1 + i g^2 C_2(G) I_{log}(\lambda^2)\,  .
\label{eqn:stir}
\ee
In other words CIR explicitly fixes the arbitrariness of IR in such a way that
gauge invariance is maintained. At higher loop order similar relations to those
displayed in equations    (\ref{CR4Q1})-(\ref{CR4L2}) are expected
to hold \cite{JHEP} and its constrained version should implement vector gauge
invariance as well \cite{WIP}.

In defining a mass-independent minimal scheme in CIR there appeared an arbitrary
non-vanishing constant $\lambda^2$. As discussed 
before, by subtracting only the term proportional to
$I_{log} (\lambda^2)$ defines a minimal subtraction scheme within IR and we are
left with the finite piece of the amplitude which is also dependent upon
$\lambda^2$. Moreover it is identical to the amplitude that we would obtain
had we employed differential renormalization \cite{PRD3}, \cite{DiffR}. The
arbitrary scales which appear in IR and Differential renormalization can be
identified and thus the truncated connected  $n$-point renormalized Green's
function, say $G^{(n)}_c (p_i,g,m)$ is  expected to satisfy a Callan-Symanzik
like renormalization group equation where $\lambda$ plays the role of
renormalization group scale. Thus we have
\bq
&& \Bigg( \lambda \frac{\partial}{\partial \lambda} + \beta(g)
\frac{\partial}{\partial g} - \gamma_m (g) m \frac{\partial}{\partial m}
\nonumber \\
&& -n_A \gamma_A (g) - n_f   \gamma_\psi (g) \Bigg)  G^{(n)}_c =0 \, ,
\eq
where $n_A$ ($n_f$) is  the number of gluon (quark)  legs in momentum space,
$m$ and $g$  are defined as in (\ref{eqn:bare}), $\gamma_A (\gamma_\psi)$ are
the anomalous dimension of the gluon (quark) field and
\bq
\beta (g) &=&  \lambda \frac{\partial g}{\partial \lambda} \, ,\nonumber \\
\gamma_m (g) &=& -\frac{\lambda}{m} \frac{\partial m}{\partial \lambda} \, ,
\nonumber \\
\gamma_A (g) &=&  \frac{\lambda}{2 Z_3} \frac{\partial Z_3}{\partial \lambda}
\,\,\,  \mbox{and} \nonumber \\
\gamma_\psi (g) &=&  \frac{\lambda}{2 Z_2} \frac{\partial Z_2}{\partial \lambda}
\, .
\eq
For instance let us explicitly calculate the $\beta$ function. Recall
(\ref{eqn:bare}): $g_0 = Z_g g$, $Z_g = Z_1 Z_3^{-3/2}$. Hence
\be
2 \lambda^2 \frac{\partial}{\partial \lambda^2} \Big( Z_g g \Big) =0  \, .
\ee
Now using that
\be
\lambda^2 \frac{\partial}{\partial \lambda^2} I_{log} (\lambda^2) = - b
\ee
in the equation above yields after some simple algebra
\be
\beta = - \frac{g^3}{3 (4 \pi)^2} \Big( 11 C_2(G) - 4 C(r) n_f \Big) + O(g^5) \,
.
\ee
In a similar fashion we may use the renormalization constants which we have
calculated in CIR to show that
\be
\gamma_m =  \frac{ 6 g^2}{ (4 \pi)^2} C_2 (r) + O(g^5) \, ,
\ee
\be
\gamma_A =  - \frac{g^2}{3  (4 \pi)^2}  \Big( 5 C_2(G) - 4 C(r) n_f \Big)+
O(g^5) \, ,
\ee
and
\be
\gamma_\psi =  \frac{g^2}{ (4 \pi)^2} C_2 (r) + O(g^5) \, ,
\ee
which are the standard values  of the renormalization group functions.
Particularly in a minimal scheme within our framework, they coincide
with the MS scheme in dimensional regularization.

\section{Conclusions}

We have applied the IR method to QCD at the one loop level. We have shown
that it preserves non abelian gauge symmetry and that there is no need of a
different prescription to deal with massive or massless theories. A constrained
version of IR which implies in momentum routing invariance also delivers gauge
invariant amplitudes for the non-abelian case.

\section*{Appendix A: }

Consider a bubble (or a piece of a certain QCD amplitude) :
\bq
I^{d}&=&\Lambda^{4-d}\int_k^d \frac{1}{(k^2-\mu^2)[(k+p)^2-\mu^2]}.
\eq
The index $d$ stands for dimensional regularization. Although IR does not use a
explicit regulator, we will use dimensional regularization here for pedagogical
purposes to show how to define a irreducible loop integral  in a massless theory
free of infrared divergencies through equation (\ref{eqn:Ilog}). Similar
relations can be derived at higher loop order \cite{JHEP}.

We  follow the prescription of IR to separate the divergent
parts by means of the identity given by equation  (\ref{eqn:rr}). As we have
discussed we may strictly assume an implicit regulator to manipulate
algebraically the integrand. However as we do not actually evaluate the
irreducible loop integrals we need not to make a regulator explicit.

\bq
I^d&=&\Lambda^{2\epsilon}\left\{\int_k \frac{1}{(k^2-\mu^2)^2}\right. \nonumber
\\ &-&\left. \int_k \frac{p^2+2p \cdot k}{(k^2-\mu^2)^2[(k+p)^2-\mu^2]}\right\},
\eq
with $\epsilon=2-d/2$. As a matter of illustration we calculate the first
integral ($I_{log}(m^2)$) using dimensional regularization to obtain
\be
\Lambda^{2\epsilon}I_{log}(m^2)=b\left[ \frac {1}{\epsilon} +A +
\ln{\left(-\frac{4\Lambda^2}{m^2}\right)}\right]+{\cal O}(\epsilon),
\ee
where $A$ is a constant characteristic of dimensional regularization.
The second integral is finite and evaluates to
\be
I^d_{fin}=-b \int_0^1\mbox{dz} \ln{\left(\frac{p^2z(1-z)-\mu^2}{-\mu^2}\right)}.
\ee
In the limit where  $\mu^2 \to 0$, we have
\be
I^d_{fin}=-b\ln{\left(-\frac{p^2}{e^2 \mu^2}\right)}.
\ee
In IR we write
\be
I^d = I_{log} (\mu^2)  -b\ln{\left(-\frac{p^2}{e^2 \mu^2}\right)}.
\ee
However because
\be
\Lambda^{2\epsilon}I_{log}(\lambda^2)=b\left[ \frac {1}{\epsilon} +A +
\ln{\left(-\frac{4\Lambda^2}{\lambda^2}\right)}\right]+{\cal O}(\epsilon),
\ee
$\lambda \ne 0$ we have
\be
\Lambda^{2\epsilon}I_{log}(\mu^2)-\Lambda^{2\epsilon}I_{log}(\lambda^2)=
b\ln{\left(\frac{\lambda^2}{\mu^2}\right)}.
\ee
which is just equation (\ref{eqn:Ilog}) in the limit $\epsilon \rightarrow 0$.
Substituting  this relation in the expression for $I^d$ yields \be
I^d=I_{log}(\lambda^2)-b\ln{\left(\frac{p^2}{e^2\lambda^2}\right)}.
\ee
Now we are allowed to subtract a genuine ultraviolet divergent object $I_{log}
(\lambda^2)$ by defining the appropriate counterterm. The non-vanishing
arbitrary parameter $\lambda$ plays the role of renormalization group scale.

\section*{Appendix B: Infrared Finiteness of the One-Loop Amplitudes}
\label{app:Antonio}

We now turn ourselves to an important discussion on 
a problem which arises when the limit $\mu^2 \to 0$ is taken. 
We must make sure that, 
when using the scale relation given by equation (7), the term in
$\ln{(\mu^2/\lambda^2)}$ will be cancelled out by a contribution that 
comes from the ultraviolet finite part. The divergent integrals that are 
present in the calculations at the one loop level for the renormalization of QCD are
\bq
A&=&\int_k \frac{1}{(k^2-\mu^2)[(k+p)^2-\mu^2]};
\\
A_{\mu}&=&\int_k \frac{k_{\mu}}{(k^2-\mu^2)[(k+p)^2-\mu^2]};
\\
B_{\mu \nu}&=& 
\int_k \frac{k_\mu k_\nu}
{(k^2-\mu^2)[(k+p)^2-\mu^2][(k+q)^2-\mu^2]} \\
B_{\mu \nu \alpha}&=& 
\int_k \frac{k_\mu k_\nu k_\alpha}
{(k^2-\mu^2)[(k+p)^2-\mu^2][(k+q)^2-\mu^2]}.
\eq
In the integrals above we have introduced the mass $\mu$, that  will be set
to zero in the end. 
Since the integrals are assumed to be regularized, we can follow the prescription 
of
Implicit Regularization, and separate the divergent parts by means of the identity 
given by equation (1). 
Below we show the calculations:
\begin{itemize}
\item $A$: 
\bq
A&=&\int_k \frac{1}{(k^2-\mu^2)^2} \nonumber \\
&-&\int_k \frac{p^2+2p \cdot k}{(k^2-\mu^2)^2[(k+p)^2-\mu^2]}.
\eq
We now use the identity expressed by equation (7)
in the first integral.
The second one, which is finite and does not depend on any specific 
technique, is given by
\be
A_{fin}=b\int_0^1\mbox{dz} \ln{\left(\frac{p^2z(1-z)-\mu^2}{-\mu^2}\right)}.
\ee
For $\mu^2 \to 0$, we have 
\be
A_{fin}=b\ln{\left(-\frac{p^2}{e^2\mu^2}\right)}.
\ee
We clearly see that the $\mu^2$ cancels out when the two parts are put together,
so that we obtain
\be
A=I_{log}(\lambda^2)-b\ln{\left(-\frac{p^2}{e^2\lambda^2}\right)}.
\ee
\item $A_\mu$: After the expansion, some integrals vanish, since their integrands
are odd in the integration variable. After calculating the finite part, we have
\bq
A_\mu=-2p^\nu \int  \frac{k_\mu k_\nu}{(k^2-\mu^2)^3}
+\frac b2 p_\mu \ln{\left(-\frac{p^2}{e^2\mu^2}\right)}.
\eq
The divergent integral is $\Theta ^{(0)}_{\mu \nu}(\mu^2)$ (see eq. (6))
and we use equation (3)
to write
\bq
\Theta ^{(0)}_{\mu \nu}(\mu^2)&=&\frac{g_{\mu \nu}}{4}\left(I_{log}(\mu^2)-\alpha_2
\right) \nonumber \\
&=&\frac{g_{\mu \nu}}{4}\left(I_{log}(\lambda^2)+b\ln{\left(\frac{\lambda^2}
{\mu^2}\right)} -\alpha_2
\right) 
\label{cons1}
\eq
Again we see the cancellation of $\mu^2$ when the finite part is considered. We are
left with
\be
A_\mu=-\frac{p_\mu}{2}\left(
I_{log}(\lambda^2)-b\ln{\left(-\frac{p^2}{e^2\lambda^2}\right)}-\alpha_2\right).
\ee
It is important to note that equation (3) is essential in the cancelation of the 
$\mu^2$.
\item $B_{\mu \nu}$: After the expansion and calculation of the finite part, we obtain
\bq
B_{\mu \nu}&=& \int_k \frac{k_\mu k_\nu}{(k^2-\mu^2)^3}
-b \frac {g_{\mu \nu}}{4} \ln{\left(-\frac{p^2}{e^2\mu^2}\right)}\nonumber \\
&+& b \left\{ \left[\left(\frac 12 +\mu^2 \eta_{00}\right)-\frac {q^2}{2} \eta_{10}
-\frac {p^2}{2} \eta_{01}\right] \right. \nonumber \\ 
&+&\left.  p_\mu p_\nu \eta_{02}+q_\mu q_\nu \eta_{20}
+(p_\mu q_\nu +q_\mu p_\nu)\eta_{11}\right\},
\eq
where
\be
\eta_{nm}=\int_0^1 \mbox{dz}\int_0^{1-z}\mbox{dy} \frac{z^n y^m}{Q(p,q,y,z)}
\ee
and
\bq
Q(p,q,y,z)&=&p^2y(1-y)+q^2z(1-z) \nonumber \\
&-&2(p\cdot q)yz-\mu^2.
\eq
It can be easily seen that the functions $\eta_{nm}$ do not have problems when 
$\mu^2\to 0$. The divergent integral has the result of equation (\ref{cons1}), so 
that
\bq
B_{\mu \nu}&=&\frac{g_{\mu \nu}}{4}\left(I_{log}(\lambda^2) 
-b\ln{\left(-\frac{p^2}{e^2\lambda^2}\right)}-\alpha_2\right) \nonumber \\
&+& f(\eta_{nm}),
\eq
where $f(\eta_{nm})$ represents the $\eta$ dependent part. 
\item $B_{\mu \nu \alpha}$: The mechanism is the same as in the other integrals: 
\bq
B_{\mu \nu \alpha}&=& -2(p+q)^\beta
\int_k \frac{k_\mu k_\nu k_\alpha k_\beta}{(k^2-\mu^2)^4}\nonumber \\
&+&\frac {b}{12}\left[(p+q)_\mu g_{\nu \alpha}
+(p+q)_\nu g_{\mu \alpha} \right.\nonumber \\
&+& \left.
(p+q)_\alpha g_{\mu \nu}\right]\ln{\left(-\frac{p^2}{e^2\mu^2}\right)}
+g(\eta_{nm}).
\eq
The $g(\eta)$ represents the $\eta$ dependent part. The divergent integral is
$\Theta^{(0)}_{\mu \nu \alpha \beta}(\mu^2)$, as defined in equation (6). 
We use the relation given by equation (5) to write 
\bq
\Theta^{(0)}_{\mu \nu \alpha \beta}(\mu^2)&=&\frac {1}{24}g_{\{\mu \nu}
g_{\alpha \beta \}}
\left(I_{log}(\mu^2)-\alpha_4\right) \nonumber \\
&=& \frac {1}{24}g_{\{\mu \nu}
g_{\alpha \beta \}}
\left(I_{log}(\lambda^2) \right. \nonumber \\
&+& \left.b\ln{\left(\frac{\lambda^2}{\mu^2}\right)} -\alpha_4\right) .
\eq
The substitution of this result in $B_{\mu \nu \alpha}$ and the adoption 
of the same procedures as before, lead us to the cancellation of the infrared 
divergences and we have the final expression
\bq
B_{\mu \nu \alpha}&=&-\frac {1}{12} 
\left[(p+q)_\mu g_{\nu \alpha}+(p+q)_\nu g_{\mu \alpha}\right. \nonumber \\
&+&\left.(p+q)_\alpha g_{\mu \nu}\right]
\left(
I_{log}(\lambda^2) \right. \nonumber \\
&-&\left. b \ln{\left(-\frac{p^2}{e^2\mu^2}\right)}-\alpha_4\right)
+ g(\eta_{nm}).
\eq
\end{itemize}
We call the reader's attention to the fact that, in the last calculation, the 
equation (5) 
was mandatory in order to $\mu^2$ to be cancelled. It is also interesting to note that 
the same relations that are necessary to preserve gauge invariance are also 
essential for 
the cancellation of the infrared cut-off $\mu^2$.


\end{document}